\documentclass[preprint,proceedings]{rmaa}

\def\etal{{et\,al.}}
\def\msun{M$_{\odot}$}
\def\rsun{R$_{\odot}$}
\def\degs{\ifmmode ^{\circ}\else$^{\circ}$\fi}

\SetYear{2004}
\SetConfTitle{Compact Binaries in the Galaxy and Beyond}

\title{Resonant Scattering and Recombination in CAL 87}

\author{
  J. Greiner\altaffilmark{1},
  A. Iyudin\altaffilmark{1,2},
  M. Jimenez-Garate\altaffilmark{3},
  V. Burwitz\altaffilmark{1},
  R. Schwarz\altaffilmark{4},
  R. DiStefano\altaffilmark{5},
  and N. Schulz\altaffilmark{3}}

\altaffiltext{1}{MPE Garching, 85741 Garching, Germany.}
\altaffiltext{2}{Moscow, Russia.}
\altaffiltext{3}{MIT, Cambridge, MA 02139, USA.}
\altaffiltext{4}{Universit\"at G\"ottingen, 37083 Germany.}
\altaffiltext{5}{CfA, Cambridge, MA 02138, USA.}

\shortauthor{Greiner et al.}
\shorttitle{CAL 87}

\fulladdresses{
\item Vadim Burwitz, Jochen Greiner and Anatoli Iyudin: 
  Max-Planck-Institut f\"ur extraterrestrische Physik,
                 Giessenbachstra\ss{}e, 85748 Garching, Germany
  (\email{burwitz, jcg, ani@mpe.mpg.de}).
\item Rosanne Di\,Stefano: Center for Astrophysics, 
  Cambridge, MA 02138, U.S.A.
  (\email{rd@cfa.harvard.edu}).
\item Mario Jimenez-Garate and Norbert Schulz: MIT, Cambridge, U.S.A.
  (\email{mario, nss@space.mit.edu}).
\item Robert Schwarz: Universit\"at G\"ottingen, 37083 G\"ottingen,
   Germany, (\email{rsc@uni-sw.gwdg.de}).
}

\listofauthors{J. Greiner, A. Iyudin, M. Jimenez-Garate, V. Burwitz,
  R. Schwarz, R. DiStefano and N. Schulz}
\indexauthor{Greiner, J.}
\indexauthor{Iyudin, A.}
\indexauthor{Jimenez-Garate, M.}
\indexauthor{Burwitz, V.}
\indexauthor{Schwarz, R.}
\indexauthor{DiStefano, R.}
\indexauthor{Schulz, N.}

\abstract{The eclipsing supersoft X-ray binary CAL 87 has been observed 
with Chandra
on August 13/14, 2001 for nearly 100 ksec, covering two full orbital cycles
and three eclipses. The shape of the eclipse light curve derived from
the zeroth-order photons indicates that the size of the X-ray emission region
is about 1.5 \rsun. The ACIS/LETG spectrum is completely dominated by
emission lines without any noticeable continuum.
The brightest emission lines are significantly redshifted and double-peaked,
suggestive of emanating in a 2000 km/s wind.
We model the X-ray spectrum by a mixture of recombination and
resonant scattering. This allows us to deduce the temperature and luminosity
of the ionizing source to be $kT \sim 50-100$ eV and
$L_X \sim 5 \times 10^{37}$ erg/s.
}

\addkeyword{binaries: close}
\addkeyword{Stars: individual: CAL 87}
\addkeyword{X-ray: stars}

\begin{document}
\maketitle

\section{Introduction}
\label{sec:intro}

CAL 87 was detected with the IPC onboard the Einstein X-ray observatory 
during the Columbia Astrophysical Laboratory survey of LMC (Long \etal\ 1981),
and it was optically identified with an eclipsing binary in the Large
Magellanic Cloud with an orbital period of 10.6 hrs
(Pakull \etal\ 1988, Callanan \etal\ 1989, Cowley \etal\ 1990).
A shallow X-ray eclipse was discovered with ROSAT (Schmidtke \etal\ 1993).

Early attempts to model the optical light curve of CAL 87 by Callanan \etal\
(1989) already indicated an elevated accretion disk rim.
This model has been extended by Schandl \etal\ (1997), and successfully 
describes the optical light curve of the CAL 87 binary system, composed of a
primary (WD) with a mass of M$_1$ = 0.75 M$_{\odot}$, placed at a distance
of 2.2$\times$10$^{11}$ cm from the mass-donating secondary star with a
mass of M$_2$= 1.5 M$_{\odot}$.
The model of Schandl \etal\ (1997) includes
(1) optical emission from the secondary star which is irradiated by the 
emission of the WD and 
(2) emission of an accretion disk with a thick rim and an optically
thick, cold, clumpy spray produced by the high mass-flow rate of the accretion
stream impinging on the disk (hot spot). This spray, moving around the disk,
nicely reproduced the asymmetry in the optical light curve and the depth of the
secondary dip.

Based on the pre-Chandra, low-resolution X-ray spectra and 
the detection of emission up to 1 keV,
CAL 87 has long been considered as one of the two hottest known SSS,
thus spurring observations with all X-ray satellites since ROSAT
(Parmar \etal\ 1997, Asai \etal\ 1998, Dotani \etal\ 2000,  
Ebisawa \etal\ 2001).

We have exploited the unique spectral capability of 
the Chandra low-energy transmission grating (LETG) 
to study in detail the X-ray emission of the canonical 
supersoft source CAL 87.

\section{Chandra observation in 2001}

We observed CAL 87 with Chandra for 97 ksec on August 13/14, 2001, using
the low-energy transmission grating (LETG/ACIS). This observation covered
three consecutive eclipses of the binary system. The phase-folded
light curve of the zeroth order photons (Fig. 1) shows a wide eclipse
with triangular shape, about $\pm$0.15 phase units wide (30\% of the orbit), 
no flat bottom and an eclipse depth of about 50\%. This suggests that
the observed X-ray emission does not come from a point source 
(e.g. the H-burning white dwarf). Instead, the occulter and the eclipsed
X-ray emission region should have about equal size. Given the orbital
period and assuming that the donor star is filling its Roche lobe (1.5 \rsun),
the size of the emission region should be $\sim$1.5 \rsun\ (at a binary 
separation of 3.5 \rsun).

\begin{figure}[t]
 \includegraphics[width=0.75\columnwidth,angle=90]{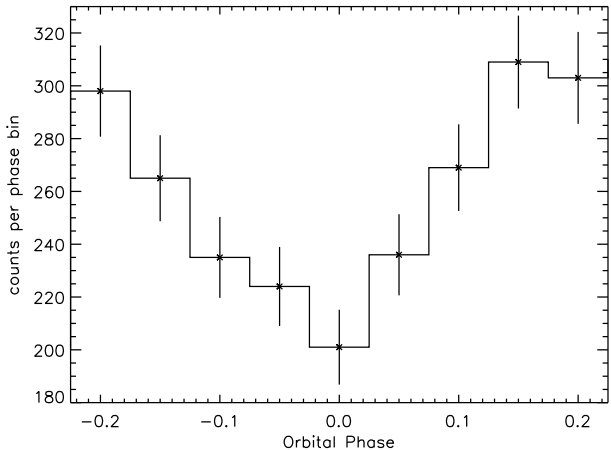}
 \caption{Phase-folded light curve of the zeroth order events around
    the eclipse minimum, showing the broad width and triangular shape
    of the eclipse.}
 \label{fig:ecl}
\end{figure}

The spectrum of CAL 87 is completely dominated by emission lines (Fig. 2), 
predominantly from oxygen, nitrogen and iron. Hardly any continuum 
emission is detectable. Galactic foreground and LMC-intrinsic absorption
cuts off the spectrum above $\sim$25 \AA. The extinction-corrected
X-ray luminosity is $\sim$10$^{35}$ erg/s.
No emission is detected shortward
of 14 \AA (corresponding to $>$0.9 keV), suggesting that the effective
temperature of the primary (expected to be thermal) X-ray emission near 
the accreting object is smaller than $\sim$ 80 eV. 

One example for the plasma diagnostic constraints which can be derived
 is the OVII line at 21.7 \AA: the strong 
resonance line and the absence of the forbidden and intercombination
lines indicates either
collisional excitation or resonance scattering. 
The prominence of the 17.05/17.10 \AA\ lines compared to the
other Fe XVII lines argues for a significant contribution
from recombination emission.
Also, the presence
of higher Ly series emission as well as the lack of the Fe XVII 15.01 \AA\
and 16.78 \AA\ emission lines (as compared to the strong 
Fe XVII 17.05/17.10 \AA\ lines) argue against collisional excitation.
Thus, as also supported by other line ratios, the X-ray spectrum of CAL 87
is mixture of recombination emission and resonant scattering.

\begin{figure}[th] 
 \includegraphics[angle=-90,width=1.\columnwidth, bb=28mm 15mm 200mm 247mm, clip]{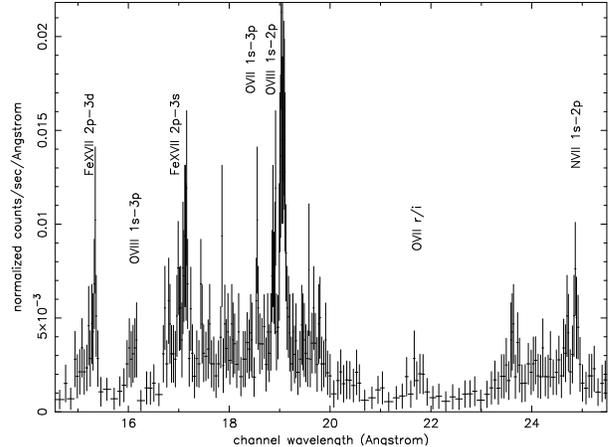} 
 \caption{Chandra LETG/ACIS count spectrum of CAL 87 summed over the bright 
     phase (excluding the eclipse). Identifications of strong emission  
     lines are given.    } 
  \label{fig:spec} 
\end{figure}

We have begun modelling the X-ray emission of CAL 87 by computing 
the emergent spectrum produced by a central source of $kT$ = 65 eV, 
$L_{\rm X}$ = 5$\times$10$^{37}$ erg/s, ionizing and being scattered off 
a corona with an electron density of 10$^{12}$ cm$^{-3}$ and log $\xi$ = 3.8 
(see Fig. 3). While the observed global features can be reproduced, 
a higher fraction of scattering is needed.  
 A first examination of the parameter space indicates 
that this can be achieved with either a softer ionizing continuum
and/or a lower column density. 

\begin{figure}[!t]  
 \includegraphics[width=1.\columnwidth,bb=35mm 145mm 190mm 252mm, clip]{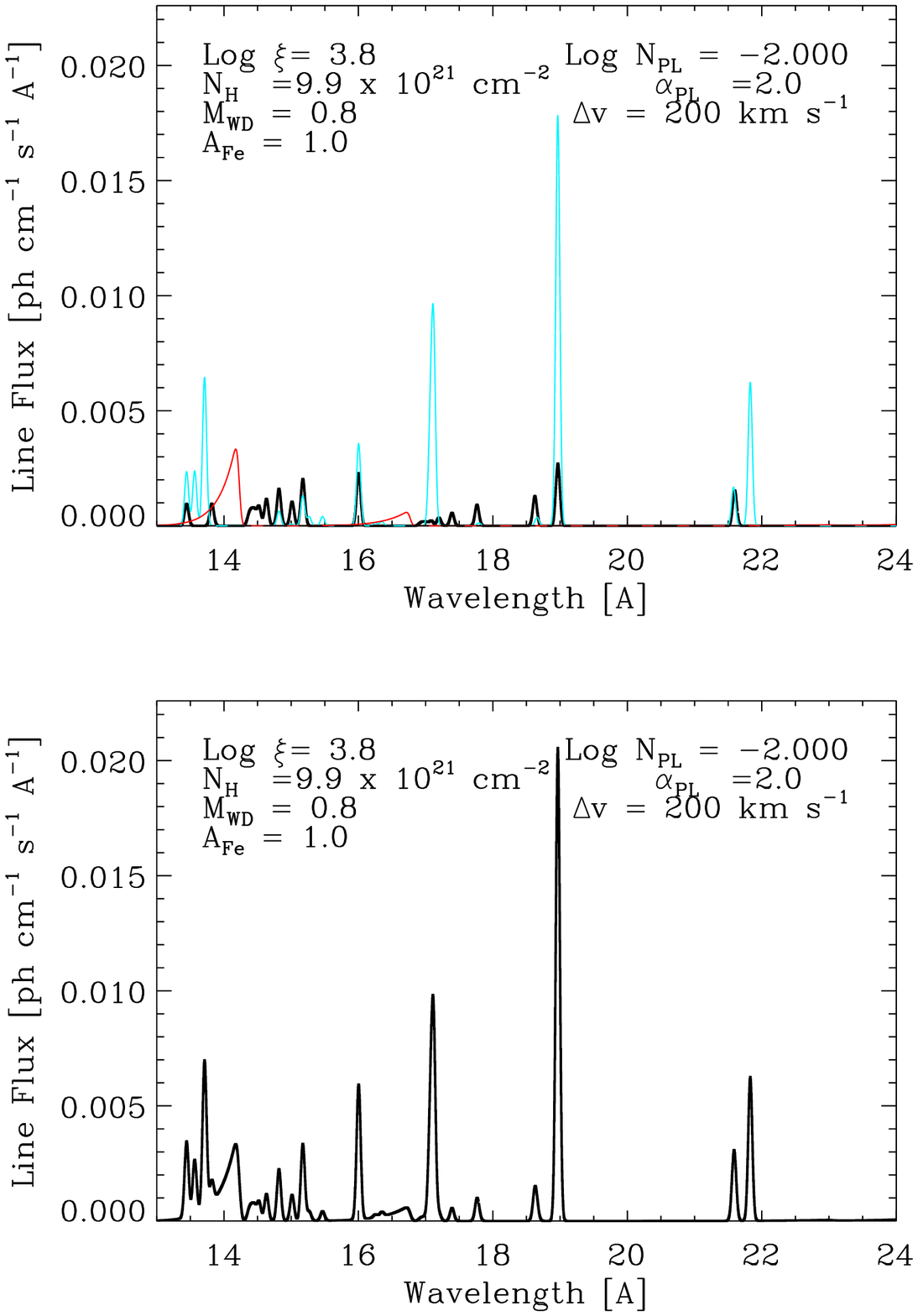}  
 \includegraphics[width=1.\columnwidth,bb=35mm 145mm 190mm 252mm, clip]{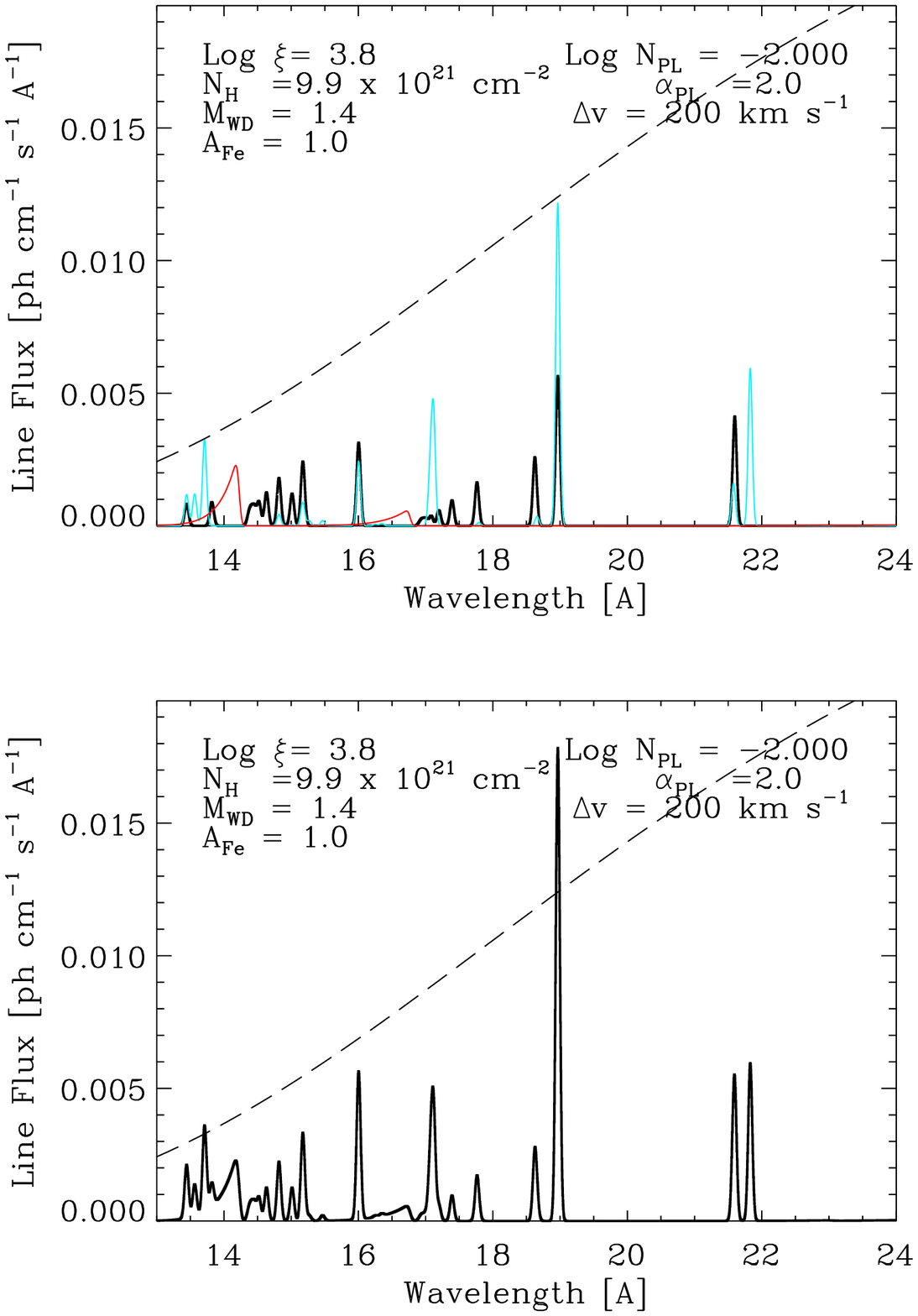}
 \vspace*{-0.7cm}
 \caption{Preliminary model of the X-ray spectrum of CAL 87, composed of
    recombination (blue) and resonant scattering (black) with
    a white dwarf mass of 0.8 \msun\ (top) and 1.4 \msun\ (bottom),
    corresponding to different effective temperatures of the ionizing source.
    The dashed line in the lower panel is the primary white dwarf spectrum.
       }  
  \label{fig:model}  
\end{figure}

An interesting property of the brightest emission line,
O VIII Lya (18.97 \AA), is that it is redshifted and has a double-peaked 
profile. The same behaviour is also noticable in the next-brightest
lines (N VII, O VII), though with lower significance.
The shift in wavelength is constant, and also does not vary with orbital phase.
In addition, the red peak disappears during eclipse.
We interpret this as a wind/outflow with an observed velocity of 1200 and
2200 km/s, respectively, after taking into account the systemic LMC velocity.
The disappearance of the red peak can be understood if the
wind/outflow is not spherically symmetric, but a bi-directional cone.
If we assume intrinsic velocity symmetry of both sides of the cone,
and de-project for the inclination of $\sim$78\degs, we obtain
an opening angle of the conical wind/outflow of  120\degs.
With such a geometry, the northern/upper part of the wind/outflow is
predominantly blue-shifted, and the southern/lower part red-shifted.
During eclipse, only the southern part of the wind/outflow
will be occulted by the donor, as observed.

\section{Conclusions}

The shape of the eclipse light curve, the lack of any continuum emission
and the dominance of emission lines due to resonant scattering and
recombination suggests that CAL 87's X-ray emission comes from an
extended wind (or outflow), 
similar to the accretion-disk-corona low-mass X-ray binary 4U 1822-37,
as already suggested by Schmidtke (1993). 
However, the plasma in 4U 1822-37 (i) has a much higher ionization
as evidenced by the H- and He-like ions of Ne, Mg, Si, S and Fe
which are completely absent in CAL 87 and (ii) is much more dominated
by recombination, indicative of larger column densities than those in CAL 87.

\medskip
\outputfulladdresses

The wind is probably not spherical, but rather has a conical shape
with an opening angle of 120\degs, extending on both sides of the 
accretion disk.
During eclipse, the donor  occults the wind on the southern disk side.


The primary X-ray emission region in CAL 87 certainly has a temperature 
substantially 
lower than $\sim$1 keV  due to the lack of Fe L, Si XIII or Ne IX/Ne X lines.
This rules out a neutron star or black hole (except for retrograde rotation)
accretor, since they would exhibit a maximum temperature of the inner part 
of the accretion disk of 0.9--1.2 keV. 
Also, the outflow
velocity of $\sim$2000 km/s is consistent with a white dwarf primary.

A corona with an ionization parameter and electron density ranging
from $>$3000/10$^{12}$ cm$^{-3}$ near the accretor 
to $\sim$30/10$^{11}$ cm$^{-3}$ near the outer boundary 
would scatter about 1\% of the
irradiated X-ray emission. This suggests that the intrinsic
X-ray luminosity of CAL 87 is of order a few times 10$^{37}$ erg/s.

Though the earlier SSS classification was based on a wrong interpretation of 
the X-ray spectrum, the new kT and L estimates from our LETG spectrum
re-confirm the supersoft source nature of CAL 87.

\acknowledgements

JG and RD gratefully acknowledge the {\it Chandra} Award No. GO1-2022X.

\end{document}